\documentclass[runningheads]{llncs}

\usepackage[T1]{fontenc}
\usepackage{graphicx}
\usepackage{hyperref} 
\usepackage{url} 
\usepackage{booktabs}
\usepackage{amsmath}
\usepackage{amsfonts}
\usepackage{nicefrac}
\usepackage{microtype}
\usepackage{lipsum}
\usepackage{fancyhdr} 
\usepackage{graphicx}
\usepackage{multirow}
\usepackage{xcolor} 
\usepackage{verbatim} 
\usepackage{soul}
\usepackage{xcolor}
\usepackage{comment}
\usepackage{algorithm}
\begin{document}

\title{Comparing ImageNet Pre-training with Digital Pathology Foundation Models for Whole Slide Image-Based Survival Analysis}

\titlerunning{Comparing ImageNet and Pathology Models for WSI Survival Analysis}

%%%%%%%%%%%%%%%%%%%%%%%%%%%%%%%%%%%%%%%%%%%%%%%%%%%%%%%%%%%%%%%%%%%%%%%%%%
%%%%%%%%%%%%%%%%%%%%%%%%%%%%%%%%%%%%%%%%%%%%%%%%%%%%%%%%%%%%%%%%%%%%%%%%%%
%%%%%%%%%%%%%%%%%%%%%%%%%%%%%%%%%%%%%%%%%%%%%%%%%%%%%%%%%%%%%%%%%%%%%%%%%%

\author{
  Kleanthis Marios Papadopoulos\inst{1} \and
  Tania Stathaki\inst{1} \and
  Nazim Benzerdjeb\inst{2,} \inst{3} \and
  Panagiotis Barmpoutis\inst{1}
}

\authorrunning{K. M. Papadopoulos et al.}

\institute{
  \inst{}Imperial College London, Department of Electrical and Electronic Engineering, London, UK \\
  \email{kp4718@ic.ac.uk} \and
  \inst{}Department of Pathology, Institut de Pathologie Multisite, Groupement Hospitalier Sud, Lyon University Hospital, Pierre-Bénite, France \and
  \inst{} University of Lyon, Université Claude Bernard Lyon 1, Lyon, France
}

%%%%%%%%%%%%%%%%%%%%%%%%%%%%%%%%%%%%%%%%%%%%%%%%%%%%%%%%%%%%%%%%%%%%%%%%%%
%%%%%%%%%%%%%%%%%%%%%%%%%%%%%%%%%%%%%%%%%%%%%%%%%%%%%%%%%%%%%%%%%%%%%%%%%%
%%%%%%%%%%%%%%%%%%%%%%%%%%%%%%%%%%%%%%%%%%%%%%%%%%%%%%%%%%%%%%%%%%%%%%%%%%

\maketitle

%%%%%%%%%%%%%%%%%%%%%%%%%%%%%%%%%%%%%%%%%%%%%%%%%%%%%%%%%%%%%%%%%%%%%%%%%%
%%%%%%%%%%%%%%%%%%%%%%%%%%%%%%%%%%%%%%%%%%%%%%%%%%%%%%%%%%%%%%%%%%%%%%%%%%
%%%%%%%%%%%%%%%%%%%%%%%%%%%%%%%%%%%%%%%%%%%%%%%%%%%%%%%%%%%%%%%%%%%%%%%%%%

\begin{abstract}
The abundance of information present in Whole Slide Images (WSIs) renders them an essential tool for survival analysis. Several Multiple Instance Learning frameworks proposed for this task utilize
a ResNet50 backbone pre-trained on natural images. By leveraging recenetly released histopathological foundation models such as UNI and Hibou, the predictive prowess of existing MIL networks
can be enhanced. Furthermore, deploying an ensemble of digital pathology foundation models yields higher baseline accuracy, although the benefits appear to diminish with more complex MIL architectures. Our code will be made publicly available upon acceptance.
\end{abstract}

%%%%%%%%%%%%%%%%%%%%%%%%%%%%%%%%%%%%%%%%%%%%%%%%%%%%%%%%%%%%%%%%%%%%%%%%%%
%%%%%%%%%%%%%%%%%%%%%%%%%%%%%%%%%%%%%%%%%%%%%%%%%%%%%%%%%%%%%%%%%%%%%%%%%%
%%%%%%%%%%%%%%%%%%%%%%%%%%%%%%%%%%%%%%%%%%%%%%%%%%%%%%%%%%%%%%%%%%%%%%%%%%

\section{Introduction}
\label{sec:intro}
Whole Slide Images (WSIs) are obtained from the digitization of histopathological glass slides and are considered the main image modality in the field of digital pathology \cite{DL_histopathology}. Digital pathology employs computational techniques to analyze whole slide images (WSIs) for various tasks, including cancer classification, tumor subtyping  and survival analysis \cite{DL_histopathology}. In clinical settings, survival analysis methods estimate the time until the occurrence of a significant event, such as death or disease recurrence.  

Traditional survival analysis techniques, including Cox proportional hazards and Kaplan-Meier models, leverage handcrafted clinical features to estimate patients' Overall Survival (OS) time \cite{UsabilityQaiser2022}. In contrast, deep learning enables automated feature extraction, prompting a shift toward techniques that leverage unstructured image features. 

Zhu et al. present DeepConvSurv in \cite{Zhu}, which is the first work to exploit the high information density of WSIs by proposing a Convolutional Neural Network (CNN) architecture for OS estimation.   

Given that WSIs are gigapixel-size images, it is not feasible to directly train neural networks on them. Instead, WSIs are divided into smaller patches which can be processed individually by the network. In addition, challenges such as limited data availability and the lack of detailed annotations render Multiple Instance Learning (MIL) a popular approach for WSI-based survival analysis. A large number of works in this field use a ResNet50 backbone pre-trained on the ImageNet dataset to produce a feature map and subsequently employ a feature aggregation mechanism in accordance with the MIL framework. 

Progress in the computer vision field is recently driven by Self-Supervised Learning (SSL) on large-scale datasets. SSL operates by leveraging unlabeled data to learn robust visual representations through pretext without relying on manual annotations \cite{hendrycks2019using}. SSL models are first pre-trained on these pretext tasks to capture general features and can then be fine-tuned on downstream tasks, allowing them to effectively adapt to specific applications \cite{hendrycks2019using}. These pre-trained models are also referred to as foundation models.

This development has inspired the release of several digital pathology foundation models, which are pre-trained on a vast number of unannotated WSIs found in public and private databases. These foundation models are considered a viable alternative to the ResNet50 backbone given their pre-training on pathology datasets instead of natural images.

This manuscript aims to provide a comprehensive comparison of the performance of Multiple Instance Learning (MIL)-based survival analysis networks using ResNet50 and two popular digital pathology foundation models. Furthermore, the use of a combination of feature extractors is explored to determine their impact on the accuracy of downstream MIL networks.

%neural networks cannot be trained on WSIs but we need to use patches.
%Background WSI based survival analysis
%What is SSL and why do we use it?
%What is special about this manuscript?
%%%%%%%%%%%%%%%%%%%%%%%%%%%%%%%%%%%%%%%%%%%%%%%%%%%%%%%%%%%%%%%%%%%%%%%%%%
%%%%%%%%%%%%%%%%%%%%%%%%%%%%%%%%%%%%%%%%%%%%%%%%%%%%%%%%%%%%%%%%%%%%%%%%%%
%%%%%%%%%%%%%%%%%%%%%%%%%%%%%%%%%%%%%%%%%%%%%%%%%%%%%%%%%%%%%%%%%%%%%%%%%%

\section{Related Work}
\subsection{Multiple Instance Learning}
Multiple Instance Learning (MIL) is a Weakly Supervised Learning paradigm widely adopted in WSI workflows. In a typical binary classification problem, a sequence of images $X=\{X_{1}, X_{2},..., X_{N}\}$ are used to train a network that can predict a label $y_{k} \in \{0,1\}$, for $k=\{1,2,...,N\}$. However, in the case of MIL, each image $X_{i}$ is a bag composed of a sequence of $M$ instances denoted by $X_{i}=\{x_{i1},x_{i2},...,x_{iM}\}$ . A label is provided for the bag, but individual instances remain unlabeled. The standard MIL assumption is that individual instances are statistically independent from each other and that their ordering does not affect the MIL algorithm's operation. This problem formulation can be extended to multi-class classification and survival analysis. There exist two approaches for tackling MIL problems:
\begin{enumerate}
    \item Instance-level approach: Each instance is assigned a score by a neural network and these scores are subsequently pooled \cite{liu2012key}.  
    \item Embedding-level approach: Bag instances are converted into compact embeddings providing a matrix of embeddings for each bag. MIL pooling is subsequently used for the downstream task \cite{wang2018revisiting}. 
\end{enumerate}
The latter approach is often preferred as it avoids introducing further bias into the classification network \cite{wang2018revisiting}. Following feature extraction, a number of MIL aggregation mechanisms can be applied. Mean-pooling and max-pooling are the earliest aggregation techniques introduced, laying the groundwork for further experimentation \cite{attentionbasedmil}. Subsequent architectures such as TransMIL \cite{transmil} leverage the ViT architecture and others introduce approaches to alleviate the high computational load associated with training ViTs. Examples include S4MIL and MambaMIL\cite{mambamil}. 

%weakly supervised learning technique that is extensively used in WSI analysis
%digital pathology foundation models
\subsection{Digital Pathology Foundation Models}
SSL models are typically trained on vast unlabeled datasets using pretext tasks, enabling them to learn robust visual representations \cite{hendrycks2019using}. After this pre-training phase, these models are fine-tuned on smaller labeled datasets that have been collected for a specific task in mind \cite{hendrycks2019using}. This approach has had great success in Natural Language Processing and has now become established in Computer Vision. Examples of pretext tasks include predicting objects' rotation angle and predicting missing image patches \cite{hendrycks2019using}.

The most popular backbone choices for pretext tasks include ResNet variants and Vision Transformers (ViTs) \cite{hendrycks2019using}. Among SSL frameworks, iBOT and DINOv2 have become prominent for ViT pre-training because of their effectiveness in learning high-quality visual features. The iBOT framework employs a masked auto-encoder approach by training models to predict masked-out patches in images \cite{zhou2022ibotimagebertpretraining}. This helps the ViT backbone learn contextual relationships, a challenge typically faced by ViTs. DINOv2 utilizes a different learning strategy known as self-distillation \cite{oquab2024dinov2learningrobustvisual}. In this approach, the model learns by comparing and aligning its representations across augmented views of the same image, enabling it to capture semantic features without reliance on labeled data \cite{oquab2024dinov2learningrobustvisual}. 

The success of the DINOv2 network in particular has motivated the release of several foundation models in the digital pathology domain. Recently released cases include the UNI model \cite{UNICHEN} by Mahmood Lab and the Hibou family of models \cite{nechaev2024hiboufamilyfoundationalvision} by HistAI. The UNI model was pre-trained on the Mass-100K dataset, which consists of over 100,000 Formalin-Fixed, Paraffin-Embedded (FFPE) Hematoxylin and Eosin (H\&E) WSIs collected from Massachusetts General Hospital (MGH) and Brigham and Women’s Hospital (BWH), as well as external slides from the GTEx consortium \cite{UNICHEN}. This curated dataset does not include images from publicly available databases such as The Cancer Genome Atlas (TCGA). The pre-training process mimics the steps followed to produce DINOv2 \cite{UNICHEN}. 

DINOv2 with Registers enhances the original DINOv2 by introducing additional tokens, otherwise known as registers, to mitigate artifacts in ViT attention maps \cite{darcet2024vision}. The Hibou model family adopts this framework for model training. Similarly to UNI,  HistAI's models were pre-trained on a private dataset of approximately 1 million images \cite{nechaev2024hiboufamilyfoundationalvision}. This collection of images includes non H\&E and cytology slides \cite{nechaev2024hiboufamilyfoundationalvision}.

%Similarly, the Hibou family of models was
% a masked autoencoder approach, trains models by predicting masked-out patches in images, which helps ViTs learn contextual relationships.
%Recent works in the field of ViT pretraining for histopathology have predominantly utilized frameworks such as iBot
%[ 4] and DINOv2 [ 5 ]. The iBot framework is used by the popular open-source model Phikon [6]. As a more recent
%and advanced framework, DINOv2 has seen adoption in several notable studies, including Virchow, RudolfV, and
%Prov-Gigapath, among others [7, 8, 9, 10, 11, 12].
%iBot
%masked autoencoder
%BYOL -> DINOv2

%%%%%%%%%%%%%%%%%%%%%%%%%%%%%%%%%%%%%%%%%%%%%%%%%%%%%%%%%%%%%%%%%%%%%%%%%%
%%%%%%%%%%%%%%%%%%%%%%%%%%%%%%%%%%%%%%%%%%%%%%%%%%%%%%%%%%%%%%%%%%%%%%%%%%
%%%%%%%%%%%%%%%%%%%%%%%%%%%%%%%%%%%%%%%%%%%%%%%%%%%%%%%%%%%%%%%%%%%%%%%%%%

\section{Methodology}
\label{sec:methodology}
\begin{figure}[!ht]
    \centering
    \includegraphics[width=0.85\linewidth]{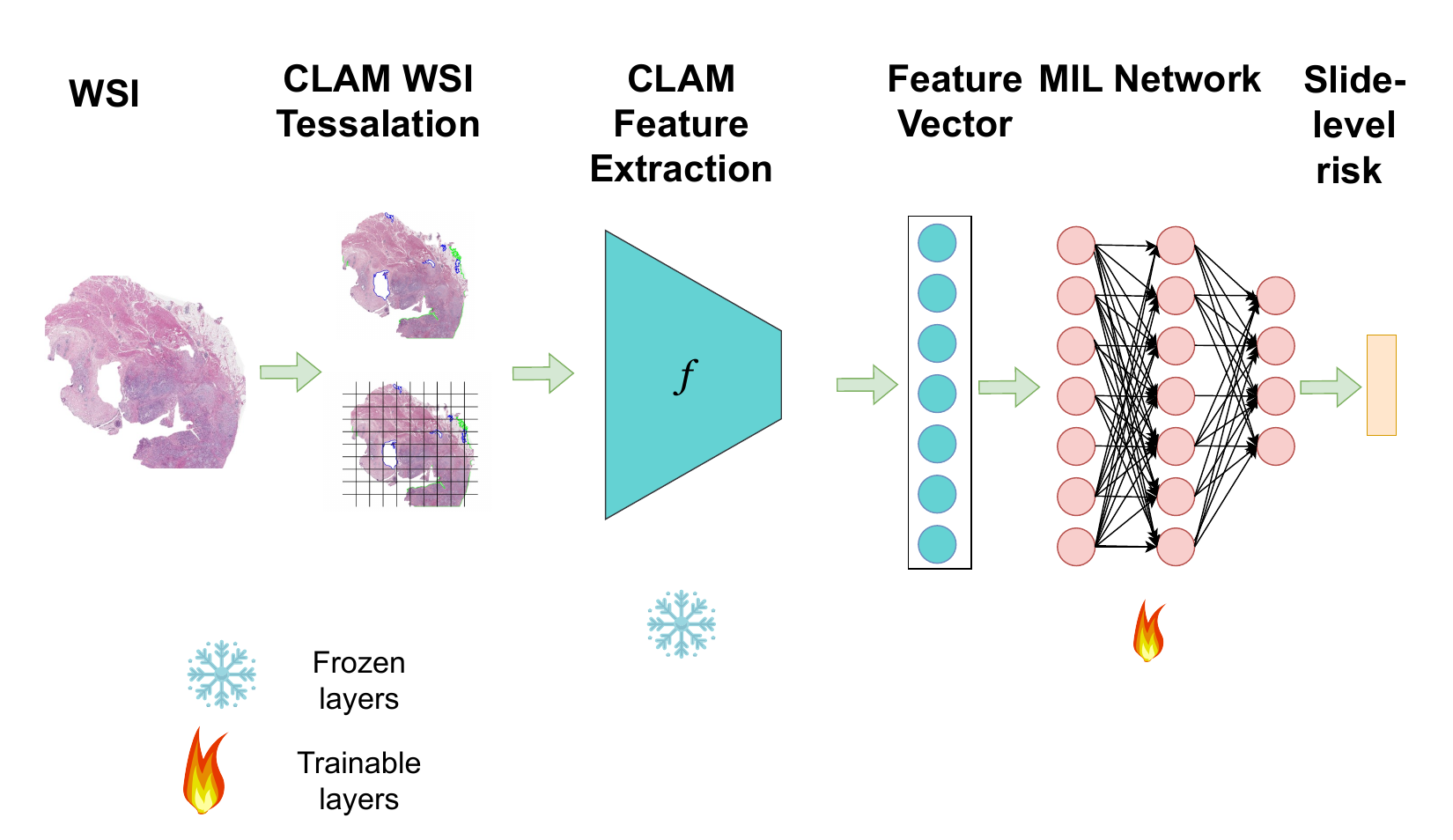}
    \caption{Overview of single backbone training strategy}
    \label{fig:nn_diagram}
\end{figure}

We tackle survival analysis as a weakly supervised slide-level task in this work and we adopt the MIL convention of fine-tuning a number of MIL frameworks following WSI feature extraction from backbone models. We utilize the Clustering-constrained Attention Multiple Instance Learning (CLAM) framework \cite{CLAM} to separate tissue content from background, segment WSIs into patches, and convert them into feature representations. We use the ResNet50 backbone as a baseline and also extract features using UNI and the Hibou-Base checkpoint made publicly available by HistAI. We subsequently train a number of downstream MIL frameworks including MeanMIL \cite{oner2020studying}, MaxMIL \cite{oner2020studying}, ABMIL \cite{attentionbasedmil}, and TransMIL \cite{transmil}. We summarize our approach in Figure \ref{fig:nn_diagram}.

In addition, we explore foundation model ensembles based on tile embeddings through feature concatenation. A feature extractor \( f_{i} \) transforms a given WSI \( X_{k} \) into a feature matrix \( \mathbf{F}_{k} \in \mathbb{R}^{m_{k} \times d} \), where $m_{k}$ denotes the number of segmented patches for the $k^{th}$ WSI and $d$ denotes the embedding space dimension. We therefore obtain a feature representation $\mathbb{G}$ mathematically defined as:
\begin{equation}
    \mathbb{G} = 
    \begin{bmatrix}
       \mathbb{F}_1 \\
       \mathbb{F}_2 \\
        \vdots \\
        \mathbb{F}_N 
    \end{bmatrix},
\end{equation}
where \( \mathbb{F}_{i} \) for \( i = 1, \dots, N \) denotes the \(i^{th} \) feature matrix.
%where \( \bold{F}_{i} \) for \( i = 1, \dots, N \) denotes the \(i^{th} \) feature extractor. 

\begin{figure}[!ht]
    \centering
    \includegraphics[width=0.8\linewidth]{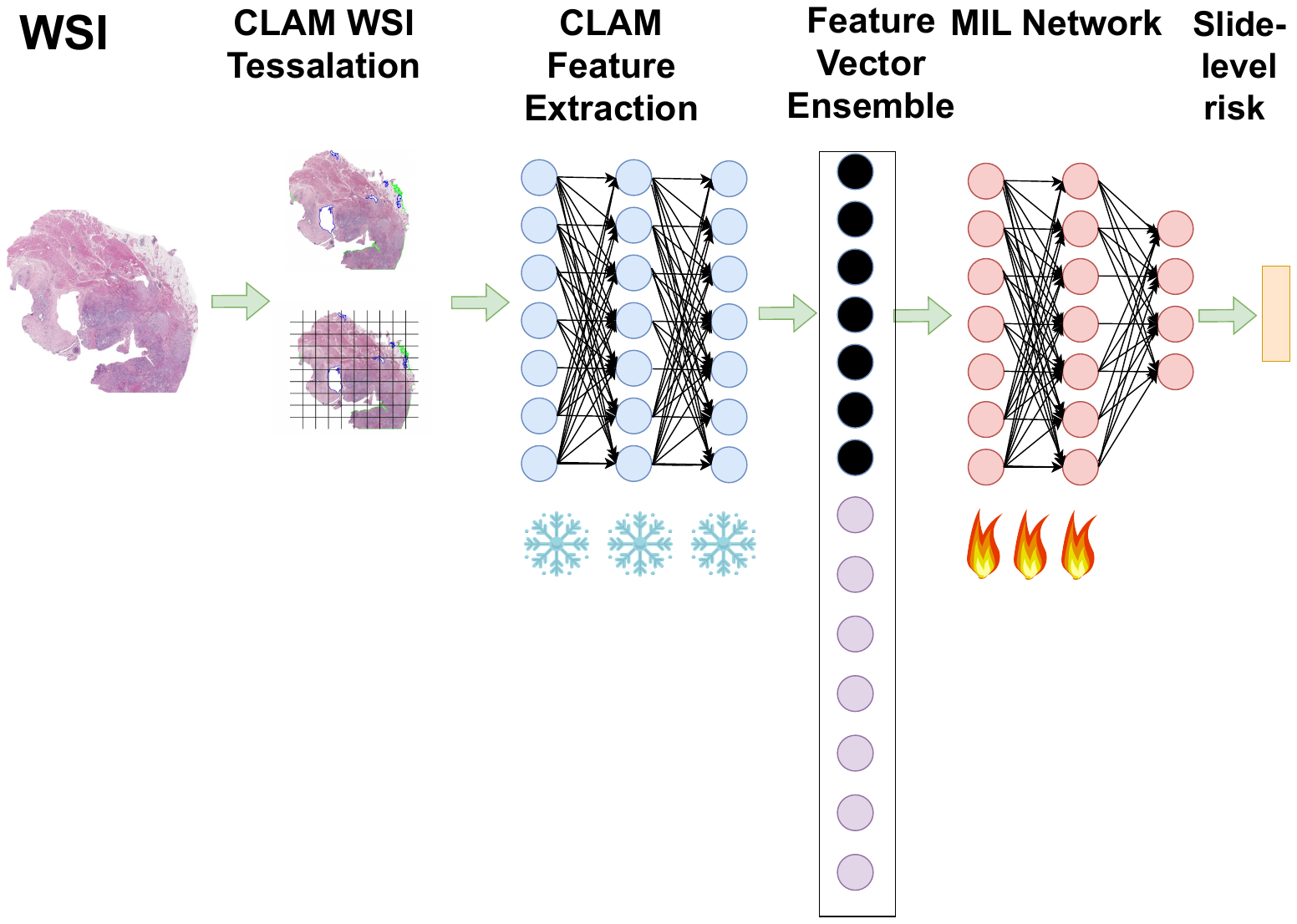}
    \caption{Overview of feature extractor ensemble training strategy}
    \label{fig:nn_ensemble_diagram}
\end{figure}

Following feature extraction, a trainable downstream MIL mechanism $h$ converts the feature map into a slide-level risk prediction $\hat{r}$:
\begin{equation}
    \hat{r}= h(\mathbb{G})
\end{equation}
We provide a graphical summary of this ensemble approach in Figure \ref{fig:nn_ensemble_diagram}. 

%we concatenate featuyres 
%we implement downstream MIl

%%%%%%%%%%%%%%%%%%%%%%%%%%%%%%%%%%%%%%%%%%%%%%%%%%%%%%%%%%%%%%%%%%%%%%%%%%
%%%%%%%%%%%%%%%%%%%%%%%%%%%%%%%%%%%%%%%%%%%%%%%%%%%%%%%%%%%%%%%%%%%%%%%%%%
%%%%%%%%%%%%%%%%%%%%%%%%%%%%%%%%%%%%%%%%%%%%%%%%%%%%%%%%%%%%%%%%%%%%%%%%%%

\section{Experiments \& Results}
\label{sec:exp}

\subsection{Datasets}
\begin{figure}[t]
    \centering
    \includegraphics[width=0.60\linewidth]{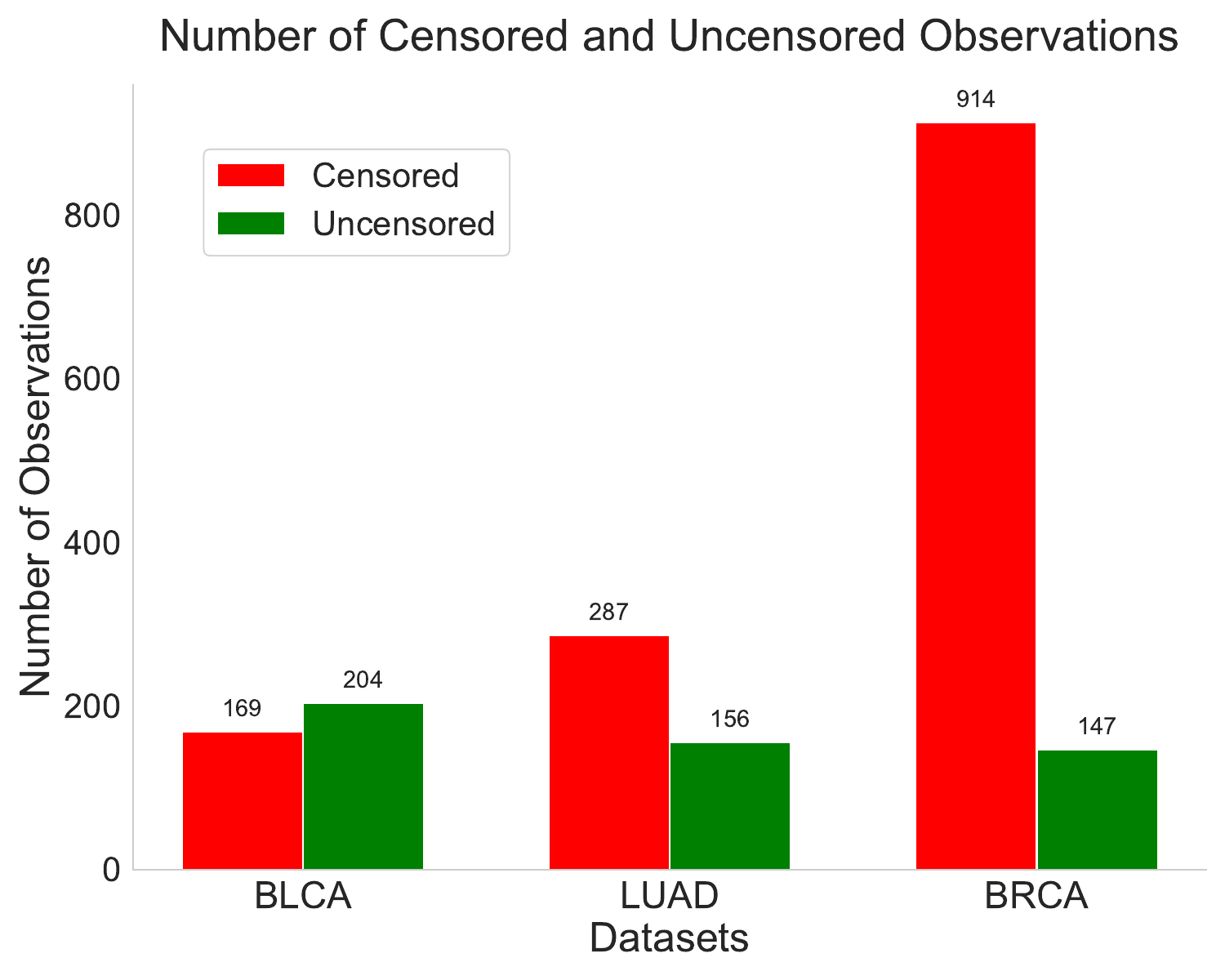}
    \caption{Distribution of censored and uncensored patients in each TCGA dataset}
    \label{fig:datasets_dist}
\end{figure}

We use WSIs from the publicly available TCGA database, and specifically the Bladder Urothelial Carcinoma (BLCA), the Lung Adenocarcinoma (LUAD), and the Breast Invasive Carcinoma (BRCA) cohorts. 
We select FFPE H\&E diagnostic slides for each cohort and we follow the convention of using one WSI per patient by selecting slides having "DX1" in their TCGA identifier. 
According to the TCGA documentation, the "DX" label denotes a diagnostic slide.  
A large number of patients have multiple "DX" slides but not all. This selection process yields $N=373$, $N=443$ and $N=1061$ WSIs for BLCA, LUAD,  and BRCA respectively. Survival analysis datasets comprise both censored and uncensored observations. Uncensored observations correspond to patients with known survival times, while censored observations pertain to patients for whom survival time is not recorded, often because the patient was alive at the time of analysis. 
Figure \ref{fig:datasets_dist} illustrates the distribution the distribution of censored and uncensored patients across each dataset, with the proportion of censored patients to the total number of patients being approximately $45\%$ for the BLCA, $65\%$ for the LUAD and $86\%$ for the BRCA dataset respectively.

\subsection{Implementation Details}
Table \ref{tab:dimensions} lists the dimension of the feature vector produced by each backbone along with the total number of parameters used during the pre-training phase. The same information for all model ensembles is provided in Table \ref{tab:embedding_dimensions}. We use the Negative Log Likelihood (NLL) as a bag-level loss function to train the aforementioned MIL networks and the concordance index to assess their prognostic accuracy. The number of trainable parameters for every MIL method is provided in Table \ref{tab:mil_models} in Appendix \ref{app: mil_model_params}. We divide the TCGA cohorts into train and validation sets and we report the results after applying $K$-fold cross validation ($K=5$).

Our codebase is inspired by the MambaMIL framework \cite{mambamil}, and we build upon its publicly available implementation. The project is implemented in PyTorch, and our code repository can be found at \url{https://github.com/MarioPaps/WSI-Survival-Analysis-ImageNet-vs-Digital-Pathology-Backbones.git}. 

For training, we use the Adam optimizer and we set the batch size equal to $1$ because the number of patches extracted for each WSI varies. We therefore perform backpropagation using gradient accumulation with the number of steps set to $32$. Furthermore, we leverage early stopping and regularization to combat overfitting.

%In terms of datasets, we use WSIs from the publicly available TCGA database. Specifically, we use the Bladder Urothelial Carcinoma (BLCA) and the Breast Invasive Carcinoma (BRCA) cohorts in our experiments. We select FFPE H\&E diagnostic slides for each cohort and we follow the convention of using one WSI per patient by selecting slides having "DX1" in their TCGA identifier. This selection process yields $N=373$ and $N=1061$ WSIs for BLCA and BRCA respectively. Survival analysis datasets comprise both censored and uncensored observations. Uncensored observations correspond to patients with known survival times, while censored observations pertain to patients for whom survival time is not recorded, often because the patient was alive at the time of analysis.

%Table \ref{tab:dimensions} lists the dimension of the feature vector produced by each backbone. We use the Negative Log Likelihood (NLL) as a bag-level loss function to train the aforementioned MIL networks and the concordance index to assess their prognostic accuracy. We divide the TCGA cohorts into train and validation sets and we report the results after applying $K$-fold cross validation ($K=5$). We list the hyperparameters used in Appendix \ref{app:hyperparams}.

\begin{table}[!ht]
\centering
\begin{tabular}{c   c   c}
\toprule
\textbf{Feature Extractor} & \textbf{Dimension} & \textbf{Number of Parameters} \\
\midrule
ResNet50 & 1024 & $25,557,032 \approx 25.6 M $ \\
UNI & 1024 & $303,350,784 \approx 303 M$  \\
Hibou-Base & 768 &  $8,5741,056 \approx 8.6 M   $ \\
\bottomrule
\end{tabular}
\vspace{1mm}
\caption{Dimensions of feature extractors used in the study}
\label{tab:dimensions}
\end{table}

\begin{table}[ht]
\centering
\begin{tabular}{c   c}
\hline
\textbf{Ensemble Combination}    & \textbf{Embedding Dimension} \\ \hline
ResNet50 + UNI             & 2048                       \\ 
Hibou-Base + ResNet50      & 1792                         \\ 
Hibou-Base + UNI           & 1792                        \\ 
ResNet50 + UNI + Hibou-Base & 2816                         \\ \hline
\end{tabular}
\vspace{1mm}
\caption{Embedding Dimensions of the ensembles}
\label{tab:embedding_dimensions}
\end{table}

\subsection{Results}
We report the $5-$fold validation results by listing the concordance index and the standard deviation for every TCGA cohort in Table \ref{table:allmethods}. We also list the concordance index values achieved by 2 model ensembles, ResNet50 with UNI and UNI with Hibou-Base, in Table \ref{table:concatmethods}. 

\begin{table} [ht]
\centering
\begin{tabular}{|l|c c c c |}
\hline
{MIL Method} & {BLCA} & {LUAD} & {BRCA} & {Average} \\ \hline
\multicolumn{5}{|c|}{\textbf{ResNet50}} \\ \hline
MeanMIL \cite{oner2020studying} & 0.536 $\pm$ 0.072 & {0.580  $\pm$ 0.021 } & 0.567 $\pm$ 0.061 & 0.561 $\pm$ 0.063 \\ \hline
MaxMIL \cite{oner2020studying} & \textbf{0.543 $\pm$ 0.044} & {0.596  $\pm$ 0.044}   & 0.570 $\pm$ 0.049 & 0.570 $\pm$ 0.056 \\ \hline
ABMIL \cite{attentionbasedmil} & 0.570 $\pm$ 0.069 & {0.594 $\pm$ 0.031} &  0.565 $\pm$ 0.040 & {0.576 $\pm$ 0.061} \\ \hline
TransMIL \cite{transmil} & 0.528 $\pm$ 0.049  & {0.597 $\pm$ 0.063}  & 0.591 $\pm$ 0.053 & 
{0.560 $\pm$ 0.051} \\ \hline

\multicolumn{5}{|c|}{\textbf{UNI}} \\  \hline
MeanMIL \cite{oner2020studying} & 0.570 $\pm$ 0.05 & {0.603  $\pm$ 0.030 } & \textbf{0.597 $\pm$ 0.085} & \textbf{0.590 $\pm$ 0.073} \\ \hline
MaxMIL \cite{oner2020studying} & 0.530 $\pm$ 0.042 & \textbf{0.625 $\pm$ 0.076} & \textbf{0.610 $\pm$ 0.092} & \textbf{0.588 $\pm$ 0.089} \\ \hline
ABMIL \cite{attentionbasedmil} & \textbf{0.601 $\pm$ 0.046} & \textbf{0.605 $\pm$ 0.061 } & {0.616 $\pm$ 0.066} & \textbf{0.607 $\pm$ 0.071} \\ \hline
TransMIL \cite{transmil} & \textbf{0.565 $\pm$ 0.014} & {0.602 $\pm$ 0.047} & \textbf{0.628 $\pm$ 0.053} & \textbf{0.598 $\pm$ 0.051} \\ \hline

\multicolumn{5}{|c|}{\textbf{Hibou-Base}} \\ \hline
MeanMIL \cite{oner2020studying} & \textbf{0.572 $\pm$ 0.036} & \textbf{0.614  $\pm$ 0.073 } & {0.575 $\pm$ 0.089} & 
{0.587   $\pm$ 0.085 } \\ \hline
MaxMIL \cite{oner2020studying} & 0.518 $\pm$ 0.031 & {0.605 $\pm$ 0.073 } & 0.550 $\pm$ 0.080 & {0.558  $\pm$ 0.080 } \\ \hline
ABMIL \cite{attentionbasedmil} & 0.570 $\pm$ 0.043 & {0.565 $\pm$ 0.080} & \textbf{0.617 $\pm$ 0.085} & {0.584  $\pm$ 0.088 } \\ \hline
TransMIL \cite{transmil} & 0.542 $\pm$ 0.053 & \textbf{0.611 $\pm$ 0.034} & {0.557 $\pm$ 0.045} & {0.570  $\pm$ 0.055 } \\ \hline
\end{tabular}

\vspace{1mm}
\caption{Concordance index values for different MIL methods across TCGA cohorts}
\label{table:allmethods}
\end{table}

\begin{table} [!ht]

\centering
\begin{tabular}{|c|   c  c  c  c|}
\hline
{MIL Method}    & {BLCA}  &  {LUAD} & {BRCA} & Average \\ \hline
      \multicolumn{5}{|c|}{\textbf{ResNet50+UNI}} \\ \hline     

      MeanMIL \cite{oner2020studying} & {0.591 $\pm$ 0.028 } & {0.586 $\pm$ 0.064}   & {0.581 $\pm$ 0.080} & 
      {0.586 $\pm$ 0.075} \\ \hline
      MaxMIL \cite{oner2020studying} &  {0.527 $\pm$ 0.050 }  & {0.603 $\pm$ 0.049} & {0.585 $\pm$ 0.089} 
      & {0.572 $\pm$ 0.080}  \\  \hline
      ABMIL \cite{attentionbasedmil} & {0.596 $\pm$ 0.034}  & {0.600 $\pm$ 0.083}
      & {0.606 $\pm$ 0.063} & {0.601 $\pm$ 0.078}
      \\  \hline
      TransMIL \cite{transmil} & \textbf{0.580 $\pm$ 0.033}  & {0.589 $\pm$ 0.050} 
      & {0.600 $\pm$ 0.078} & {0.590  $\pm$ 0.070 } \\ \hline
    %  S4-MIL\cite{s4mil} & aa & aa  \\
    %  hh \cite{yang2024mambamil} &  aa & aa \\ %submitpbs 9451890
      \multicolumn{5}{|c|}{\textbf{UNI+Hibou-Base}} \\ \hline 

      MeanMIL \cite{oner2020studying}  & \textbf{0.615 $\pm$ 0.020 }  & {0.601 $\pm$ 0.059 }  & {0.634 $\pm$ 0.068} 
      & {0.617 $\pm$ 0.065} \\ \hline
      MaxMIL \cite{oner2020studying} & \textbf{0.544 $\pm$ 0.050 }  & \textbf{0.615 $\pm$ 0.032}
      & \textbf{0.616 $\pm$ 0.069} 
      & \textbf{0.592 $\pm$ 0.064} \\  \hline
      ABMIL \cite{attentionbasedmil} & {0.587 $\pm$ 0.026}  & \textbf{0.623 $\pm$ 0.063}
      & {0.610 $\pm$ 0.008} & {0.607 $\pm$ 0.049}  \\  \hline
      TransMIL \cite{transmil} & {0.551 $\pm$ 0.025}  & {0.578 $\pm$ 0.060}
      & {0.608 $\pm$ 0.058} & {0.579 $\pm$ 0.062} \\  \hline
      %S4-MIL \cite{s4mil} & aa & aa\\
      %hh \cite{yang2024mambamil} & aa & aa  \\

      \multicolumn{5}{|c|}{\textbf{ResNet50+Hibou-Base}} \\ \hline 
      MeanMIL \cite{oner2020studying}  & {0.603 $\pm$ 0.020 }  & \textbf{0.605 $\pm$ 0.058 }   & \textbf{0.656 $\pm$ 0.080 } 
      & \textbf{0.621 $\pm$ 0.071}  \\ \hline
      MaxMIL \cite{oner2020studying} & {0.530 $\pm$ 0.034  }  & {0.592 $\pm$ 0.046} & {0.563  $\pm$ 0.064} 
      & {0.562 $\pm$ 0.061}  \\  \hline
      ABMIL \cite{attentionbasedmil} & {0.579 $\pm$ 0.034}  & {0.604 $\pm$ 0.089} & \textbf{0.655 $\pm$ 0.078} & {0.613 $\pm$ 0.087}  \\  \hline
      TransMIL \cite{transmil} & {0.561 $\pm$ 0.021} & \textbf{0.599 $\pm$ 0.050} & \textbf{0.643 $\pm$ 0.057} & {0.601 $\pm$ 0.056} \\  \hline

    \multicolumn{5}{|c|}{\textbf{ResNet50 + UNI + Hibou-Base}} \\ \hline 
      MeanMIL \cite{oner2020studying}  & {0.603 $\pm$ 0.022}  & {0.595 $\pm$ 0.079 }   & {0.650 $\pm$ 0.074}
      &  {0.616 $\pm$ 0.078}  \\ \hline
      MaxMIL \cite{oner2020studying} & {0.525 $\pm$ 0.044}  & {0.603 $\pm$ 0.049} & {0.570 $\pm$ 0.063} 
      & {0.566  $\pm$ 0.064}  \\  \hline
      ABMIL \cite{attentionbasedmil} & \textbf{0.598 $\pm$ 0.022}  & {0.604 $\pm$ 0.096} & \textbf{0.644 $\pm$ 0.058} 
      & \textbf{0.615 $\pm$ 0.081} \\  \hline
      TransMIL \cite{transmil} & {0.563 $\pm$ 0.036 }  & {0.590 $\pm$ 0.040} &  {0.609 $\pm$ 0.053} & \textbf{0.616 $\pm$ 0.078} \\  \hline

\end{tabular}
\vspace{2mm}
\caption{Concordance index values for feature ensembles of MIL methods across TCGA cohorts }
\label{table:concatmethods}
\end{table}

\section{Conclusions}
\label{sec:conclusions}
\label{sec:conclusion}
The results in Table \ref{table:allmethods} indicate that both histopathological feature extractors can consistently enhance the predictive prowess of the MIL networks used in this study. In most cases, the UNI backbone achieves a larger improvement over ResNet50 compared to Hibou-Base. 

The reported concordance index values in Table \ref{table:concatmethods} demonstrate that a significant performance improvement can be obtained by utilizing an ensemble of feature extractors. 
This improvement is particularly pronounced with the two simpler MIL architectures, which are MeanMIL \cite{oner2020studying} and MaxMIL \cite{oner2020studying}. 
Despite the ensemble of ResNet50 and UNI having a larger embedding dimension, the combinations of Hibou-Base with ResNet50 and UNI deliver more frequent improvements in outcome prediction accuracy. Furthermore, while combining two histopathological feature extractors typically results in a higher concordance index, mixing a feature extractor pre-trained on ImageNet with one pre-trained on WSIs outperforms the dual histopathological extractor approach in some cases.
Lastly, the ensemble of all three feature extractors does not appear to yield a notable performance improvement in any MIL architecture.

%The results in Table \ref{table:allmethods} indicate that both histopathological feature extractors can consistently enhance the predictive prowess of the MIL networks used in this study. In most cases, the UNI backbone achieves a larger improvement over ResNet50 compared to Hibou-Base. The concordance index values reported in Table \ref{table:concatmethods} reveal that an ensemble of the two digital pathology backbones consistently outperforms an ensemble consisting of a feature extractor pre-trained on ImageNet with one pre-trained on WSIs. Furthermore, it is worth noting the performance improvement of utilizing an ensemble of feature extractors. This enhancement is especially evident with MeanMIL, though the benefit diminishes with more complex MIL network architectures. It is also noteworthy that this improvement occurs despite the ensemble of ResNet50 and UNI having a larger embedding dimension than the UNI and Hibou-Base ensemble.

Future work includes incorporating feature extractors pre-trained using different SSL approaches, besides self-distillation employed in DINOv2, to further assess the effect of different strategies on prognostic accuracy.

%, particularly in the case of MeanMIL.
% However, this improvement is less pronounced when employing MIL networks with more complex architectures.

%Future work includes incorporating feature extractors pre-trained using different SSL approaches, besides self-distillation employed in DINOv2, to further assess the effect of different strategies on prognostic accuracy.
%SMALLER SIZE OF EMBEDDING DIM MENTION
%This happens despite the fact that the ensemble of ResNet50 and UNI has a larger embedding dimension than the ensemble of UNI and Hibou-Base
%Future work will consider feature extractors pre-trained using different SSL approaches beyond self-distillation, such as that used in DINOv2, to further explore how various strategies affect prognostic accuracy.

%%%%%%%%%%%%%%%%%%%%%%%%%%%%%%%%%%%%%%%%%%%%%%%%%%%%%%%%%%%%%%%%%%%%%%%%%%
%%%%%%%%%%%%%%%%%%%%%%%%%%%%%%%%%%%%%%%%%%%%%%%%%%%%%%%%%%%%%%%%%%%%%%%%%%
%%%%%%%%%%%%%%%%%%%%%%%%%%%%%%%%%%%%%%%%%%%%%%%%%%%%%%%%%%%%%%%%%%%%%%%%%%

\section{Acknowledgments}
We acknowledge computational resources and support provided by the Imperial College Research Computing Service (http://doi.org/10.14469/hpc/2232). 

\appendix  
\refstepcounter{section}
\section*{Appendix \Alph{section}: MIL Model Parameters}
\addcontentsline{toc}{section}{Appendix \Alph{section}: MIL Model Parameters}

%\renewcommand{\thesection}{Appendix \Alph{section}}

%\section{MIL Model Parameters}
\renewcommand{\thetable}{A.\arabic{table}} % Custom format for appendix A
\setcounter{table}{0} % Reset table counter for Appendix A
\label{app: mil_model_params}
Table \ref{tab:mil_models} presents the number of trainable parameters for MeanMIL \cite{oner2020studying}, MaxMIL \cite{oner2020studying}, ABMIL \cite{attentionbasedmil}, and TransMIL \cite{transmil}, assuming an input embedding dimension of $1024$.

\begin{table}[!ht]
    \centering
    
    \begin{tabular}{c   c}
    \hline
    \textbf{MIL Network} & \textbf{Number of Parameters} \\ \hline
        MeanMIL \cite{oner2020studying} & $526,852  \approx 527K$  \\ \hline
        MaxMIL \cite{oner2020studying}  & $526,852  \approx 527K $  \\ \hline
        ABMIL \cite{attentionbasedmil} &  $592,645  \approx 593K$ \\  \hline
        TransMIL \cite{transmil} & $2,673,172 \approx 2.7M$ \\  \hline
    \end{tabular}
 
\vspace{2mm}
    \caption{Trainable parameter counts for MIL networks with an input embedding dimension of 1024.}
       \label{tab:mil_models}
    
\end{table}
 
%%%%%%%%%%%%%%%%%%%%%%%%%%%%%%%%%%%%%%%%%%%%%%%%%%%%%%%%%%%%%%%%%%%%%%%%%%
%%%%%%%%%%%%%%%%%%%%%%%%%%%%%%%%%%%%%%%%%%%%%%%%%%%%%%%%%%%%%%%%%%%%%%%%%%
%%%%%%%%%%%%%%%%%%%%%%%%%%%%%%%%%%%%%%%%%%%%%%%%%%%%%%%%%%%%%%%%%%%%%%%%%%
\refstepcounter{section}
\section*{Appendix \Alph{section}: Hyperparameter Choices}
\addcontentsline{toc}{section}{Appendix \Alph{section}: Hyperparameter Choices}
\renewcommand{\thetable}{B.\arabic{table}} % Custom format for appendix A
\setcounter{table}{0} % Reset table counter for Appendix A
\label{app:hyperparams}
Table \ref{table:hyperparameters} summarizes the hyperparameter choices that are consistently applied across all model setups.
\begin{table}[ht]
\centering
\caption{Hyperparameter options common across all model configurations}
\vspace{2mm}
\begin{tabular}{l|c c c }
\textbf{Datasets} & \textbf{BLCA} & \textbf{LUAD} & \textbf{BRCA} \\ \hline
Number of WSIs & $373$ & $443$ & $1061$  \\ \hline
Batch size & $1$ & $1$ & $1$ \\ \hline
Gradient accumulation steps & $32$ & $32$ & $32$ \\ \hline
Optimizer & Adam & Adam & Adam \\ \hline
Bag weight & $0.7$ & $0.7$ & $0.7$ \\ \hline
Learning rate & $2e-4$ & $1e-4$ & $5e-5$ \\ \hline
Epochs & $200$ & $200$  & $200$  \\ \hline
Early stopping & Enabled & Enabled & Enabled \\ \hline
Earliest stop epoch & $40$ & $40$ & $40$ \\ \hline
L1 regularization & $1e-4$ & $1e-4$ & $1e-4$ \\ \hline
Patience epochs & $10$ & $5$ & $10$ \\ \hline
Dropout rate & 0.25 & 0.25 & 0.25 \\ \hline
Weight decay & $1e-3$ & $5e-4$ & $5e-4$  \\ \hline
\end{tabular}
\label{table:hyperparameters}
\end{table}

%%%%%%%%%%%%%%%%%%%%%%%%%%%%%%%%%%%%%%%%%%%%%%%%%%%%%%%%%%%%%%%%%%%%%%%%%%

\bibliographystyle{ieeetr}  
\bibliography{refs}  

%%%%%%%%%%%%%%%%%%%%%%%%%%%%%%%%%%%%%%%%%%%%%%%%%%%%%%%%%%%%%%%%%%%%%%%%%%
%%%%%%%%%%%%%%%%%%%%%%%%%%%%%%%%%%%%%%%%%%%%%%%%%%%%%%%%%%%%%%%%%%%%%%%%%%
%%%%%%%%%%%%%%%%%%%%%%%%%%%%%%%%%%%%%%%%%%%%%%%%%%%%%%%%%%%%%%%%%%%%%%%%%%

\end{document}